\documentclass[aps,twocolumn,superscriptaddress,noshowpacs,
noshowkeys]{revtex4-1}
\usepackage{amssymb, amsmath}
\usepackage{color}
\usepackage{graphicx}

\begin{document}
\title{Temperature induced tunable particle separation} 
\author{A. S{\l}apik}
\affiliation{Institute of Physics and Silesian Center for Education and Interdisciplinary Research, University of Silesia, 41-500 Chorz{\'o}w, Poland}
\author{J. {\L}uczka}
\affiliation{Institute of Physics and Silesian Center for Education and Interdisciplinary Research, University of Silesia, 41-500 Chorz{\'o}w, Poland}
\author{J. Spiechowicz}
\affiliation{Institute of Physics and Silesian Center for Education and Interdisciplinary Research, University of Silesia, 41-500 Chorz{\'o}w, Poland}
\affiliation{Institute of Physics, University of Augsburg, D-86135 Augsburg, Germany}
\begin{abstract}
An effective approach to isolation of sub-micro sized particles is desired to separate cancer and healthy cells or in therapy of Parkinson's and Alzheimer's disease. However, since bioparticles span a large size range comprising several orders of magnitude, development of an adequate separation method is a challenging task. We consider a collection of non-interacting Brownian particles of various sizes moving in a symmetric periodic potential and subjected to an external unbiased harmonic driving as well as a constant bias. We reveal a nonintuitive, yet efficient, separation mechanism based on {\it thermal fluctuations induced} negative mobility phenomenon in which particles of a given size move in a direction opposite to the applied bias. By changing solely temperature of the system one can separate particles of various strictly defined sizes.  This novel approach may be important step towards development of point-of-care lab-on-a-chip devices.
\end{abstract}
\maketitle
\section{Introduction}
Separation and fractionation of micro- and sub-micro-sized particles has ever growing importance in both research and industrial applications including chemical and biological research as well as medical diagnostics \cite{yager2006}. For example, detection and treatment of HIV disease relies on the isolation of human T-lymphocytes from whole blood \cite{cheng2007}. Similarly, separation of neuronal cells plays pivotal role in cell replacement therapy of neurodegenerative disorders such as Parkinson's and Alzheimer's disease \cite{korecka2007}. The bioparticle size is often a signature of abnormal biological properties leading to disease. This is apparent e.g. for mitochondria and lipid droplets where anomalous size indicates Huntington's disease \cite{heffner1978} or leukemia \cite{eguchi1987}. In some cases cancer cells are found to differ in size as compared to healthy ones \cite{suresh2007}. Therefore efficient strategies for separation of bioparticles are required in order to investigate variations of biomolecular signatures. 

Unfortunately, bioparticles span a large size range comprising several orders of magnitude starting from hundreds of nanometers to tens of micrometers \cite{bhagat2010, xuan2014}. For such sub-micro scale thermal fluctuations are lead actors and isolation techniques are rather scarce \cite{sajeesh2014, sonker2019}. An ideal solution would be a tunable method which allows to change a bioparticle size targeted for separation by controlling one of its parameters. 
In the following we demonstrate a nonintuitive, yet efficient, novel separation strategy taking advantage of a paradoxical mechanism of thermal noise induced absolute negative mobility (ANM) \cite{eichhorn2002,machura2007,spiechowicz2014pre}. 
Its main advantage is that it combines benefits of both active and passive separation techniques. The method utilizes an external driving force as well as a constant bias so the particle sorting efficiency and throughput are expected to be higher than for alternative passive and some active (e.g. ratchet) techniques. On the other hand, since the separation process is induced and controlled by thermal fluctuations it can be applied also to electrically neutral objects which carry no charge or dipole. Moreover, this scheme allows to not only deflect different particle species along different transport angles but even to steer them in opposite directions and therefore it is ideal for separation and fractionation purposes. Finally, the same setup can be applied to segregate particles  with respect to their mass  \cite{slapik2019}.

A key finding of this development is that the omnipresent thermal fluctuations are not necessarily a redundant nuisance but rather may provide novel, tunable mechanism for particle separation at the sub-micro scale. It requires only two ingredients: (i) a \emph{symmetric} spatially periodic nonlinear structure and (ii) \emph{nonequilibrium} state created by e.g. a time periodic driving force of vanishing mean value. We show that under an additional action of a constant bias,  thermal fluctuations guide particles of a given size in the direction opposite to this net force whereas the others move concurrently towards it, all of that under identical experimental conditions. Moreover, we demonstrate that by changing only temperature of the system one is able to tune the negative mobility effect solely for a  precisely defined size of the particle therefore allowing to separate it from the other of different sizes. 

Our setup can be experimentally realized using a lab-on-a-chip device consisting of microfluidic structure. The oscillating force driving the system out of equilibrium may be induced through hydrodynamic flow, but electrophoresis, electroosmosis or dielectrophoresis can also be utilized \cite{sonker2019}. In particular, the proof of principle experiment of a similar separation scheme has been already performed with insulator dielectrophoresis in a nonlinear, symmetric microfluidic structure with electrokinetically activated transport \cite{ros2005, eichhorn2010}. Recently, such setup allowed to induce ANM not only for colloidal particle but even for biological compound in the form of mouse liver mitochondrium \cite{luo2016}. However, these experiments harvested for separation purposes the deterministic ANM whereas here we present a essentially different approach based on the thermal noise induced phenomenon. Temperature driven tunability of the proposed particle separation process is possible to achieve solely for this latter fundamentally distinct mechanism of the ANM. 

\section{Model}
We study a collection of non-interacting classical  Brownian particles of various sizes which move in a spatially periodic potential \mbox{$U(x) = U(x + L)$} of period $L$ and are  additionally subjected to an unbiased time-periodic force $A\cos{(\Omega t)}$ of amplitude $A$ and angular frequency $\Omega$, as well as an external static force $F$. Dynamics of a single particle of mass $M$ is described by the following Langevin equation \cite{hanggi2009}
\begin{equation}
\label{model}
	M\ddot{x} + \Gamma\dot{x} = -U'(x) + A\cos{(\Omega t)} + F + \sqrt{2\Gamma k_B T}\,\xi(t),
\end{equation}
where the dot and the prime denote differentiation with respect to time $t$ and the particle coordinate $x$, respectively. The coupling of the particle with thermal bath of temperature $T$ is modelled by Gaussian white noise of zero mean and unity intensity, i.e., 
\begin{equation}
	\langle \xi(t) \rangle = 0, \quad \langle \xi(t) \, \xi(s) \rangle = \delta(t-s).
\end{equation}
The parameter $k_B$ is the Boltzmann constant and $\Gamma$ is the friction coefficient.  The potential $U(x)$ is assumed to be \emph{symmetric} and in the simplest form, namely, 
\begin{equation}
	U(x) = \Delta U \sin \left(\frac{2\pi}{L}x\right).
\end{equation}
Despite its apparent simplicity the studied model serves as a paradigmatic example exhibiting peculiar transport behaviour including, among others, noise enhanced transport efficiency \cite{spiechowicz2014pre,spiechowicz2016jstatmech}, anomalous diffusion \cite{spiechowicz2017scirep}, amplification of normal diffusion \cite{reimann2001, spiechowicz2015chaos, lindner2016} and the non-monotonic temperature dependence of normal diffusion \cite{spiechowicz2016njp, spiechowicz2017chaos}. 

We first recast  Eq. (\ref{model}) into its dimensionless form. This procedure ensures that the later obtained results are setup independent which is essential to facilitate the choice in realizing the best setup for testing our theory by experimentalist. 
Here, we  rescale the particle coordinate and time  as  
\begin{equation}
\label{scaling}
	\hat{x} = \frac{x}{L}, \quad \hat{t} = \frac{t}{\tau_\gamma}, \quad \tau_\gamma = L \sqrt{\frac{M}{\Delta U}}, 
\end{equation}
which transform  Eq. (\ref{model}) to the form
\begin{equation}
\label{dimlessmodel}
    \ddot{\hat{x}} + \gamma \dot{\hat{x}} = - \hat{U}'(\hat{x}) + a \cos{(\omega \hat{t})} + f + \sqrt{2 D} \, \hat{\xi}(\hat{t}).
\end{equation}
The \textit{dimensionless friction coefficient} $\gamma$ is a ratio of the two characteristic time scales, i.e.
\begin{equation}
\label{gamma}
\gamma = \frac{\tau_\gamma}{\tau_0} = 
\frac{\Gamma L}{\sqrt{M \Delta U}},
\end{equation}
where $\tau_0 = M/\Gamma$. The  parameter $\gamma$ is crucial for the proposed size-based separation because it depends, via the Stokes formula,  on the linear size $R$ of the particle.  
E.g. for a  spherical particle  $\Gamma = 6\pi \eta R$, where $\eta$ is the viscosity of the surrounding medium and $R$ is the radius of the spherical particle. 
Let us note that a (sub-)microsized particle typically possesses rather small physical mass $M$ and therefore the rescaled friction coefficient $\gamma$ is expected to be of moderate to large magnitude as compared to the dimensionless mass $m$ which in this scaling is set to unity $m = 1$. Other parameters read $a = (L/\Delta U) A$, $\omega = \tau_\gamma \Omega$, $f = (L/\Delta U) F$. The rescaled potential is $\hat{U}(\hat{x}) = U(L\hat{x})/\Delta U = \sin{(2\pi \hat{x})}$ and possesses the unit period $\hat{U}(\hat{x}) = \hat{U}(\hat{x} + 1)$. The dimensionless thermal noise $\hat{\xi}(\hat{t})$ has the same statistical properties as $\xi(t)$ and  $D = k_BT/\Delta U$  is a ratio of thermal energy to the half of non-rescaled potential barrier. 
From now on, only the dimensionless variables will be used in this study and therefore, in order to simplify the notation, the \emph{hat} symbol will be omitted.
\begin{figure*}[t]
	\centering
	\includegraphics[width=0.49\linewidth]{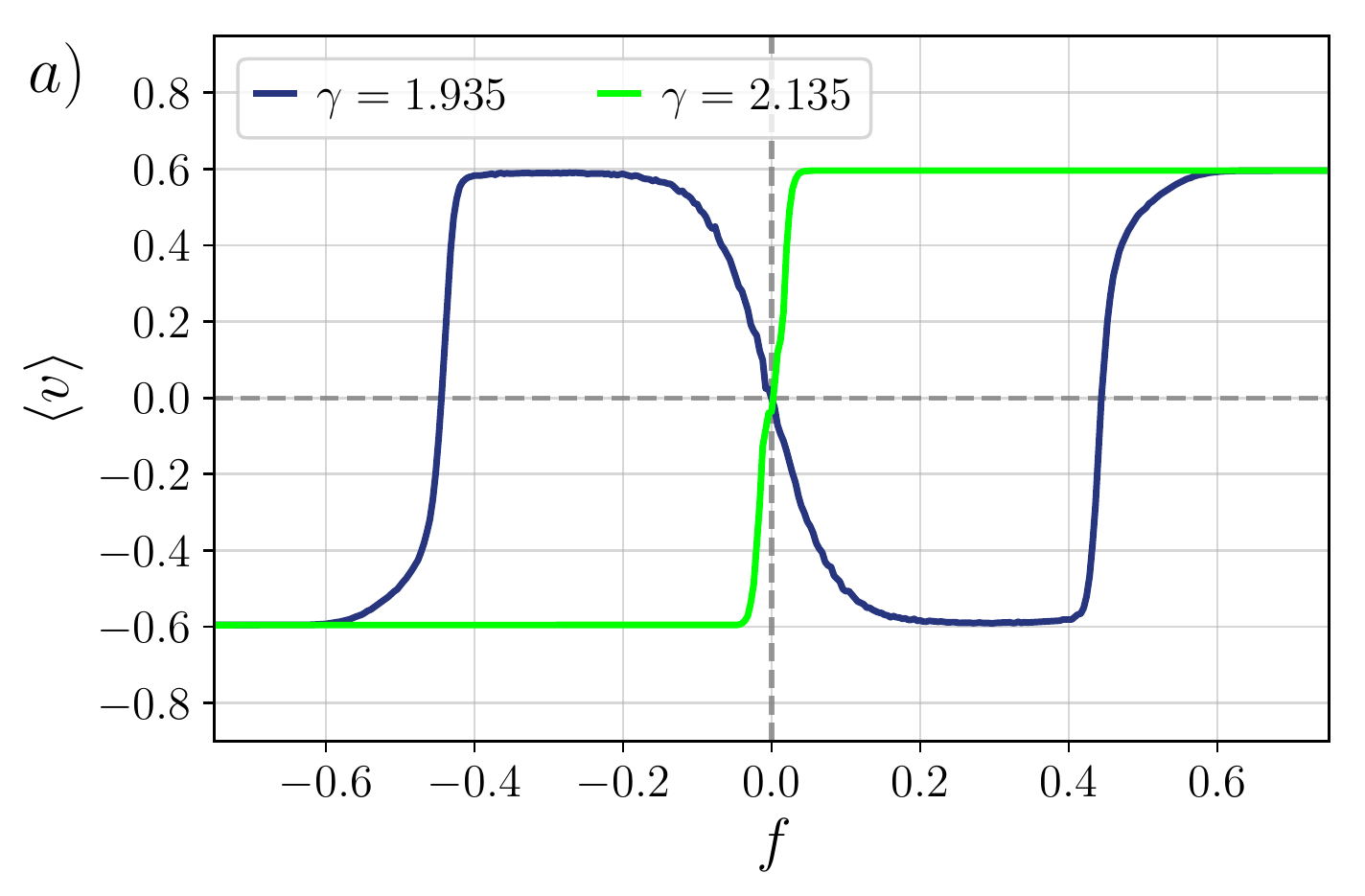}
	\includegraphics[width=0.49\linewidth]{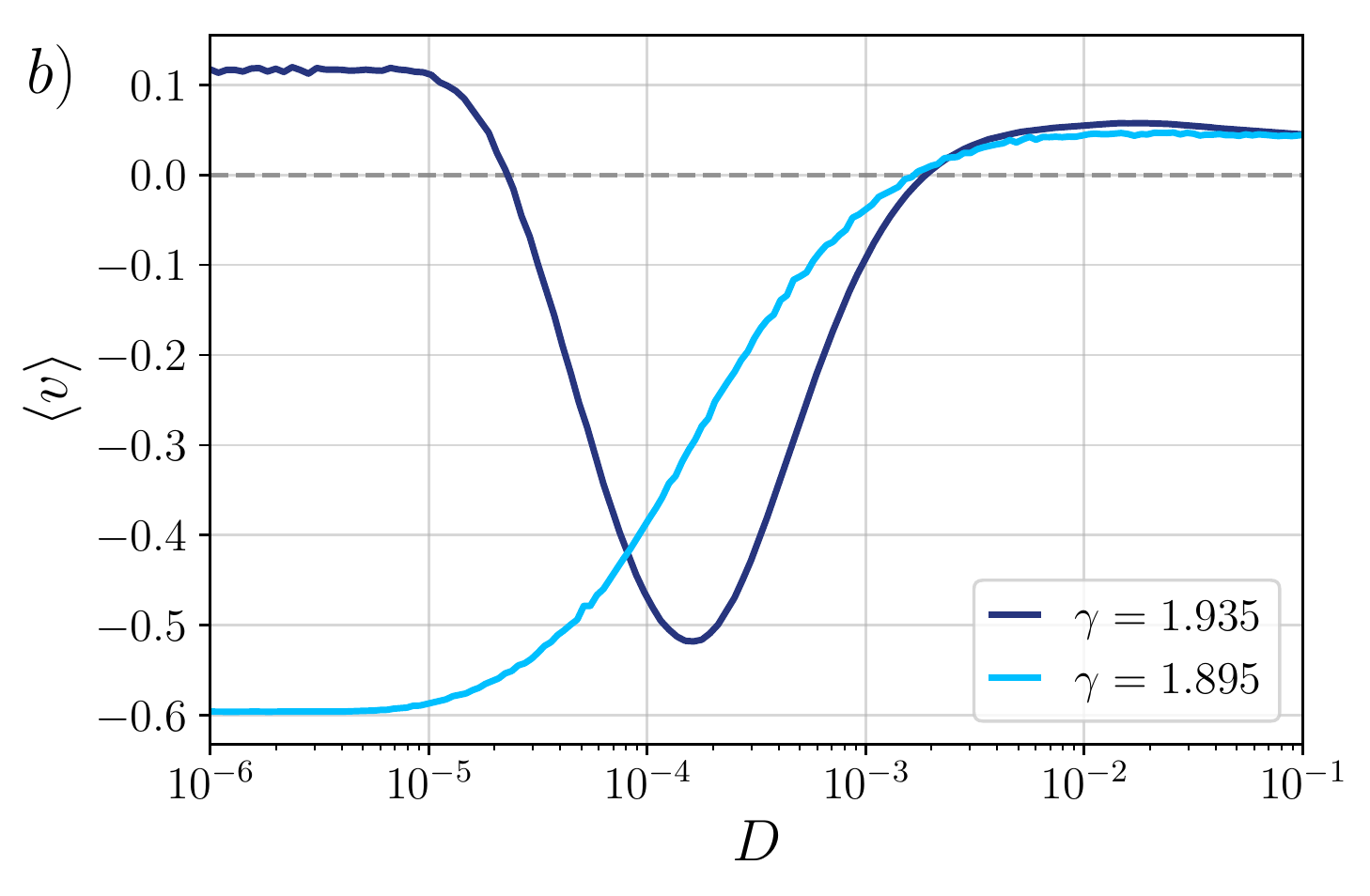}
	\caption{Panel (a): the average velocity $\langle v \rangle$ depicted as a function of the static force $f$ for two different values of the friction $\gamma$. Panel (b): the average velocity $\langle v \rangle$ presented versus thermal noise intensity $D \propto T$ for different values of $\gamma$. Other parameters read: $a = 5.75, \omega = 3.75, f = 0.1$ and $D = 0.0002$.}
	\label{fig1}
\end{figure*}

\subsection{Methods}
The observable of foremost interest in this study is the stationary averaged velocity $\langle v \rangle$ which can be expressed as
\begin{equation}
\label{directedvelocity}
\langle v \rangle = \lim_{t \to \infty} \frac{1}{t} \int_0^t ds \langle \dot{x}(s) \rangle,
\end{equation}
where $\langle \cdot \rangle$ indicates averaging over all realizations of thermal noise as well as over initial conditions for the particle position $x(0)\in[0,1]$ and its velocity $\dot{x}(0)\in[-2,2]$ uniformly distributed.  The latter is obligatory for the deterministic limit $D \propto T \to 0$  when dynamics may be non-ergodic and results can be affected by specific choice of initial conditions \cite{spiechowicz2016scirep}.

Unfortunately, the Fokker-Planck equation corresponding to the Langevin Eq. (\ref{dimlessmodel}) cannot be solved in a closed form. For this reason  we were forced to carry out comprehensive numerical simulations. Dynamics described by equation (\ref{dimlessmodel}) is characterized by a 5-dimensional parameter space $\{\gamma, a, \omega, f, D\}$, the detailed exploration of which is a very challenging task.
We carried out numerical simulations of the equation (\ref{dimlessmodel}) over a wide domains of parameters in the area $\gamma \times a \times \omega \in [0.1,10] \times [0,25] \times [0,20]$ at a resolution of 400 points per dimension, for several values of the force $f$ taken from the interval $[0,2]$ and thermal noise intensity $D \in [0, 10^{-1}]$. 
Overall, we considered nearly $10^9$ different parameter sets. This exceptional precision was made possible only thanks to our innovative simulation method which is based on employing GPU supercomputers, for details see Ref. \cite{spiechowicz2015cpc}. In particular, we employed a weak 2nd order predictor-corrector method \cite{platen} with the time step scaled by the fundamental period $\mathsf{T} = 2\pi/\omega$ of the external driving force, i.e. $h = 10^{-2} \times \mathsf{T}$. Since we are interested in the asymptotic long time state of the system numerical stability is an extremely important problem to obtain reliable results. Fortunately, predictor-corrector algorithm is similar to implicit methods but it does not require the solution of the algebraic equation at each step. It offers good numerical stability which it inherits from the implicit counterpart of its corrector. The stationary averaged velocity $\langle v \rangle$ was averaged over the ensemble of $2^{10} = 1024$ trajectories, each starting with different initial condition according to the distribution presented above. The number of realisations of stochastic dynamics is not accidental but it was so chosen to maximize the numerical simulation performance. The quantity of interest was calculated after it reached its asymptotic stationary value which typically takes place after $10^4$ periods of the external harmonic driving $\mathsf{T} = 2\pi/\omega$.

\section{Results}
Our idea how to separate sub-micro sized particles is the following. The stationary averaged velocity $\langle v \rangle$ depends, via dynamics determined by Eq. (\ref{dimlessmodel}), on the friction coefficient $\gamma = \gamma(R)$ which in turn is a function of the particle size $R$. Assume that there is a collection of e.g. four types of particles with $R_1<R_2<R_3<R_4$. We want to separate only particles of the radius $R_2$. To this aim we should find such a parameter regime $\{a, \omega, f, D\}$ in which the particles of sizes $R_1, R_3$ and $R_4$ move in the positive direction $\langle v \rangle > 0$ whereas the particles with radius $R_2$ travel in the negative direction $\langle v \rangle < 0$. Then only the particles of size $R_2$ will be separated from the rest. Let us now look for parameter regimes permitting such isolation process by harvesting the ANM phenomenon.

In the normal transport regime (outside of ANM), for sufficiently small values of the external perturbing bias $f$ the Green-Kubo linear response theory holds true and the standard response of the system is that the average particle velocity  is an increasing function of the static force, i.e. $\langle v \rangle = \mu f$. However, there are also regimes for which particles move on average in the direction \textit{opposite} to the applied bias, namely $\langle v \rangle < 0$ for $f > 0$, exhibiting anomalous behaviour in the form of \emph{negative mobility} $\mu < 0$ \cite{machura2007,speer2007,nagel2008}. The key ingredient for the occurrence of the latter effect is that the system is driven far away from thermal equilibrium into a time-dependent nonequilibrium state \cite{machura2007,hanggi2009,speer2007}. In our case this condition is guaranteed by the presence of the external harmonic driving $a\cos{(\omega t)}$.
\begin{figure*}[t]
	\centering
	\includegraphics[width=0.49\linewidth]{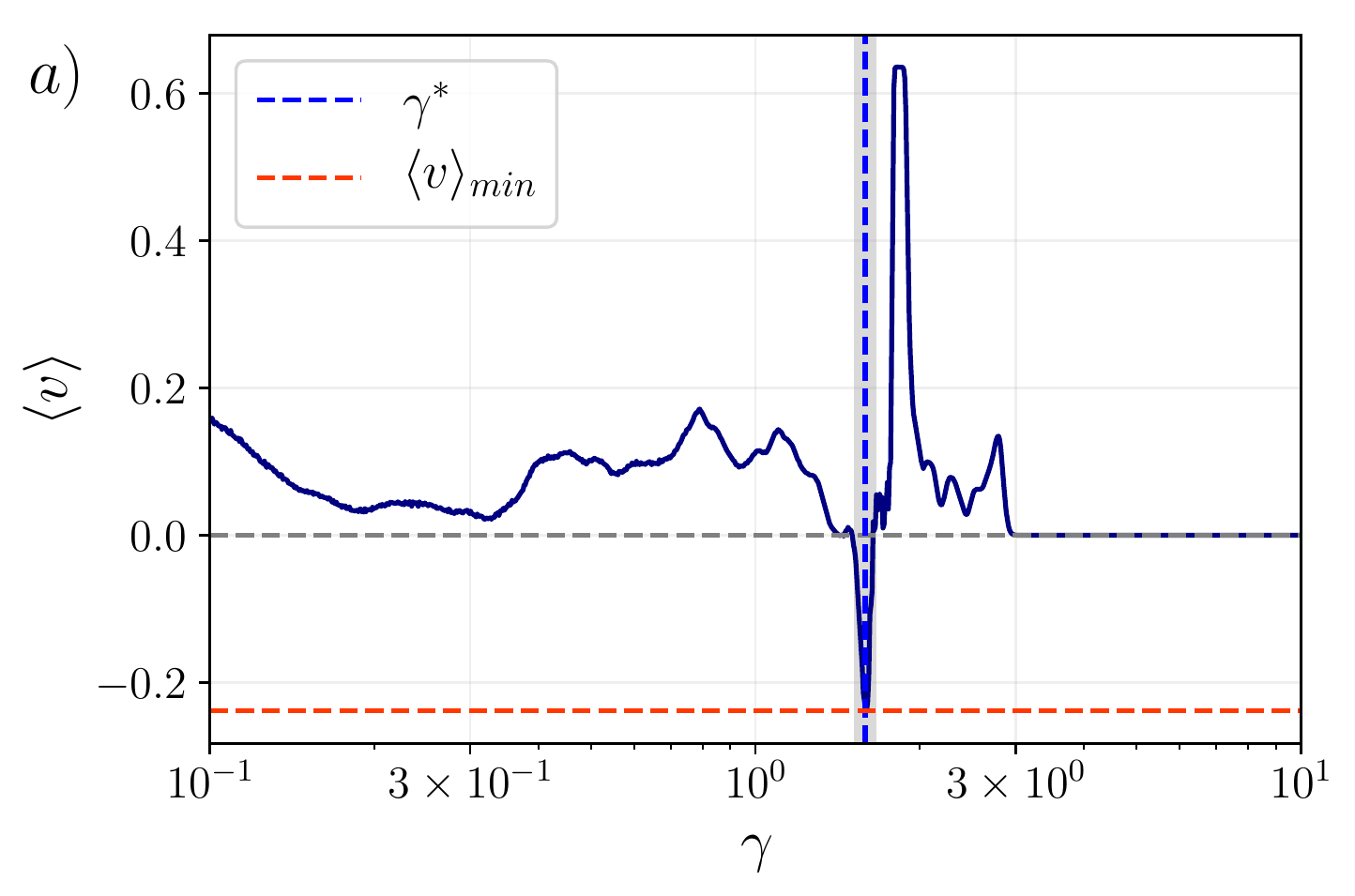}
	\includegraphics[width=0.49\linewidth]{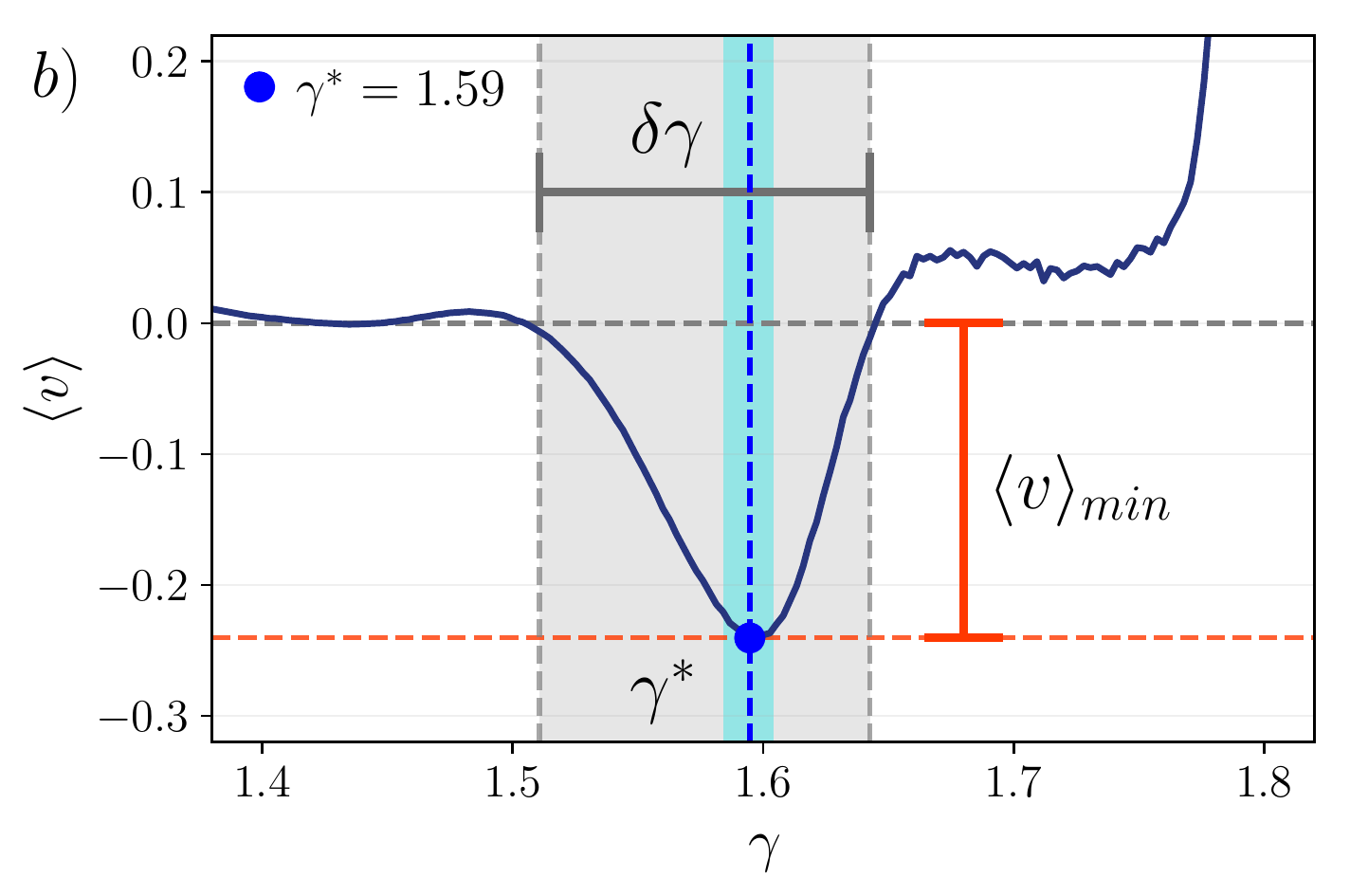}
	\caption{Average velocity $\langle v \rangle$ versus the friction coefficient $\gamma$ is depicted in the panel (a). ANM is observed only for a narrow interval marked by a grey colour. Panel (b) presents a blow up of this region. Parameter values: $a = 5.75$ $\omega = 4.0$, $f = 0.1$ and $D = 0.0006$.}
	\label{fig3}
\end{figure*}
\begin{figure*}[t]
	\centering
	\includegraphics[width=0.49\linewidth]{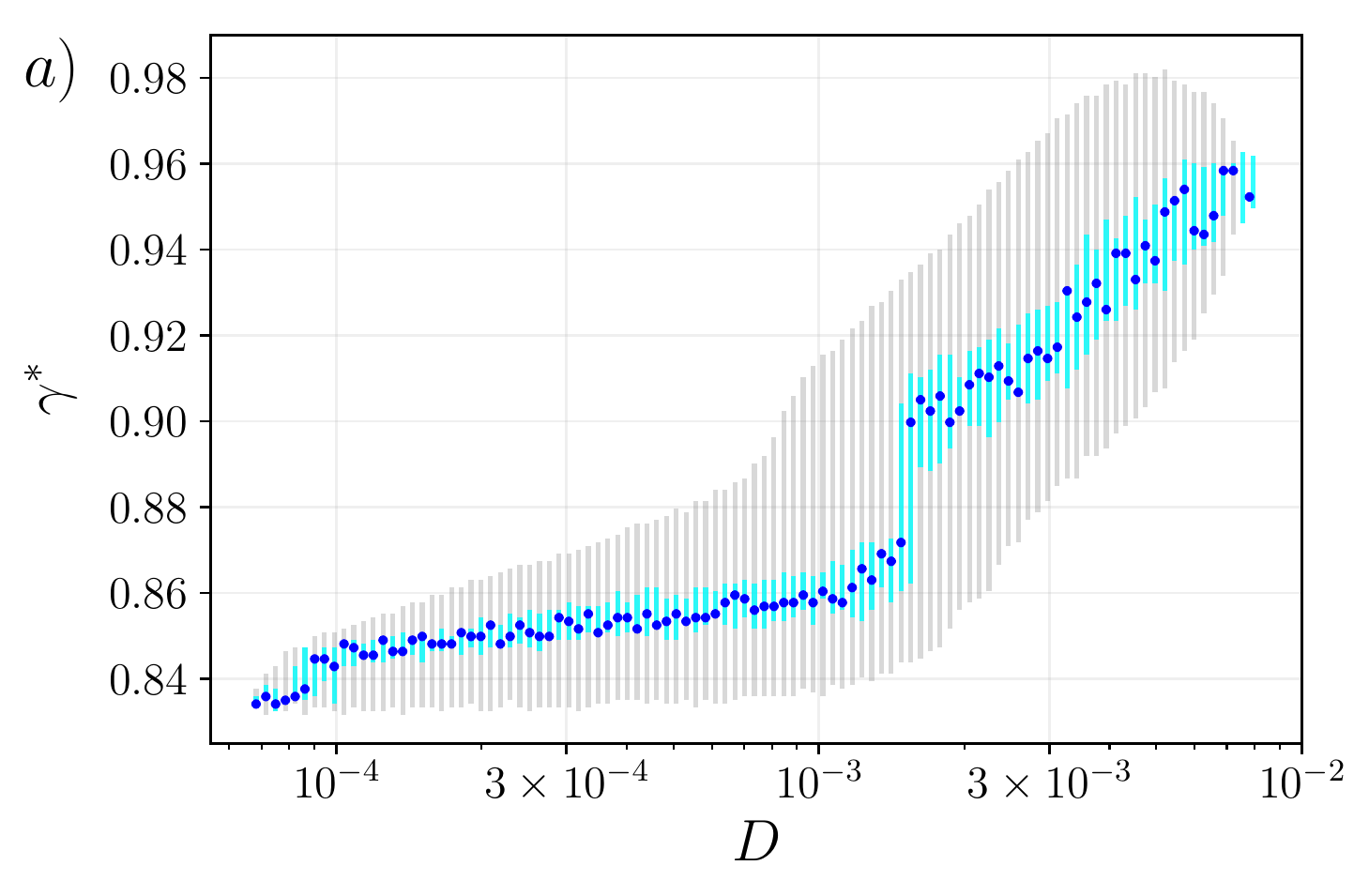}
	\includegraphics[width=0.49\linewidth]{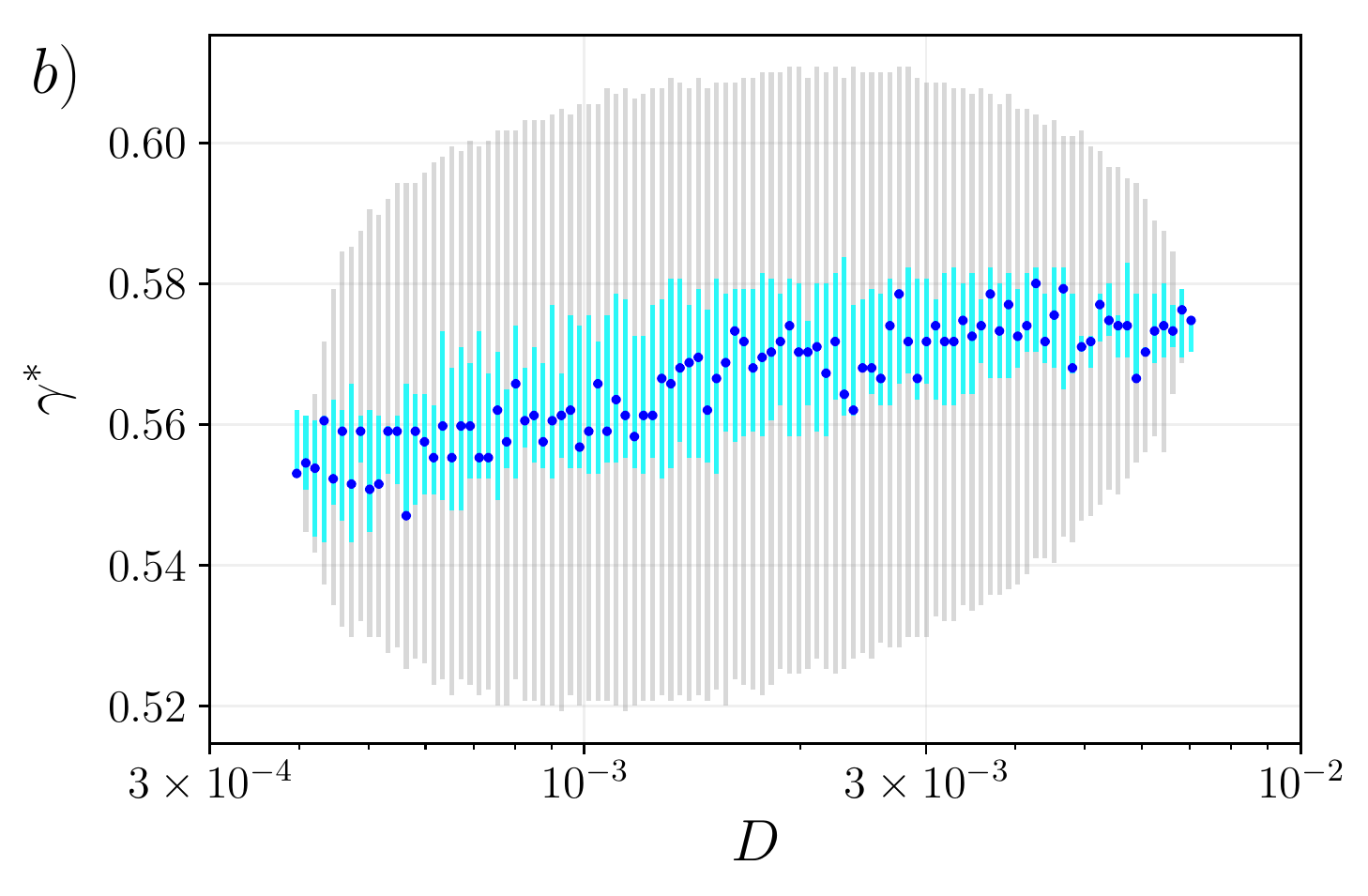}\\
	\includegraphics[width=0.49\linewidth]{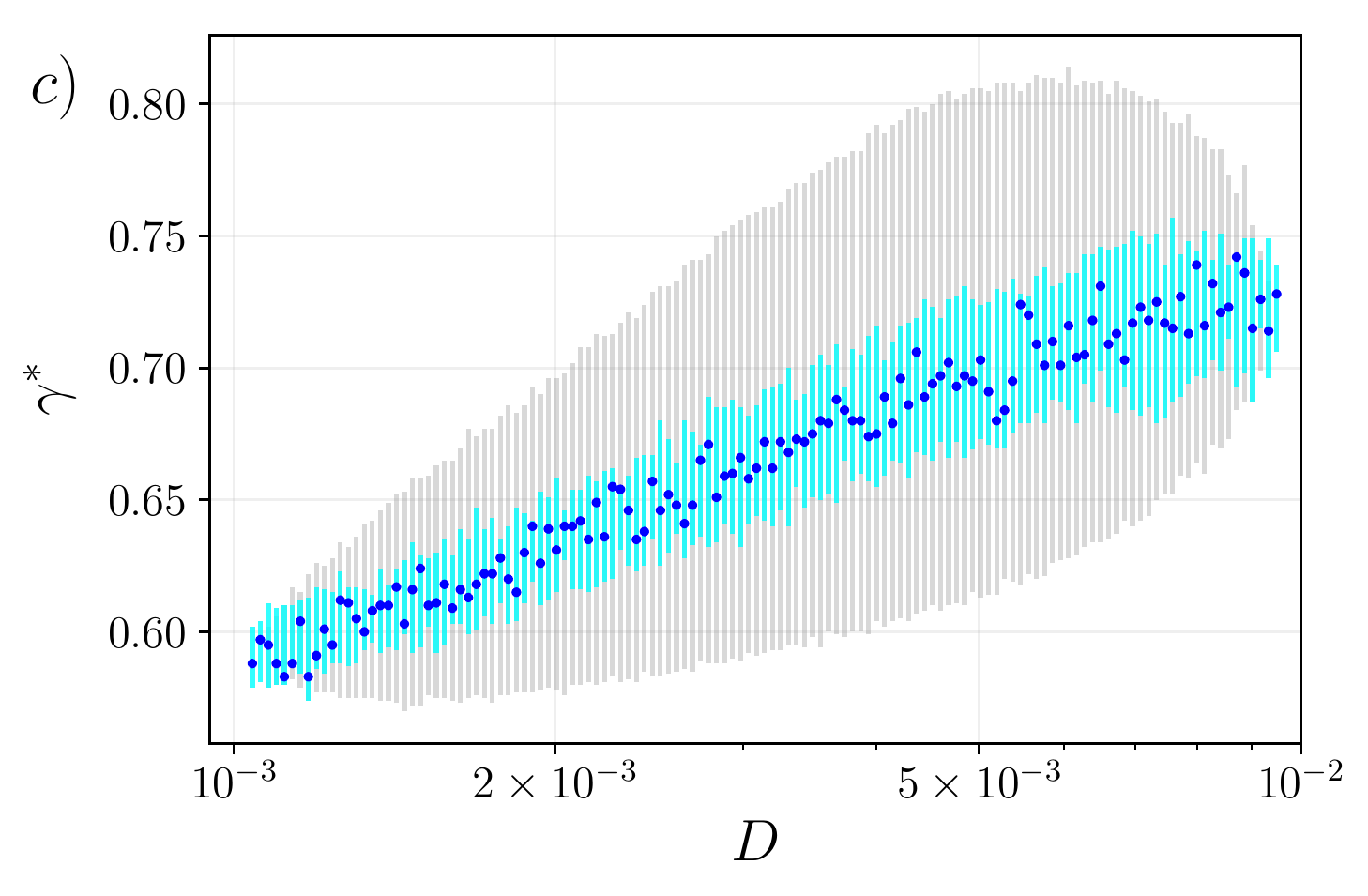}
	\includegraphics[width=0.49\linewidth]{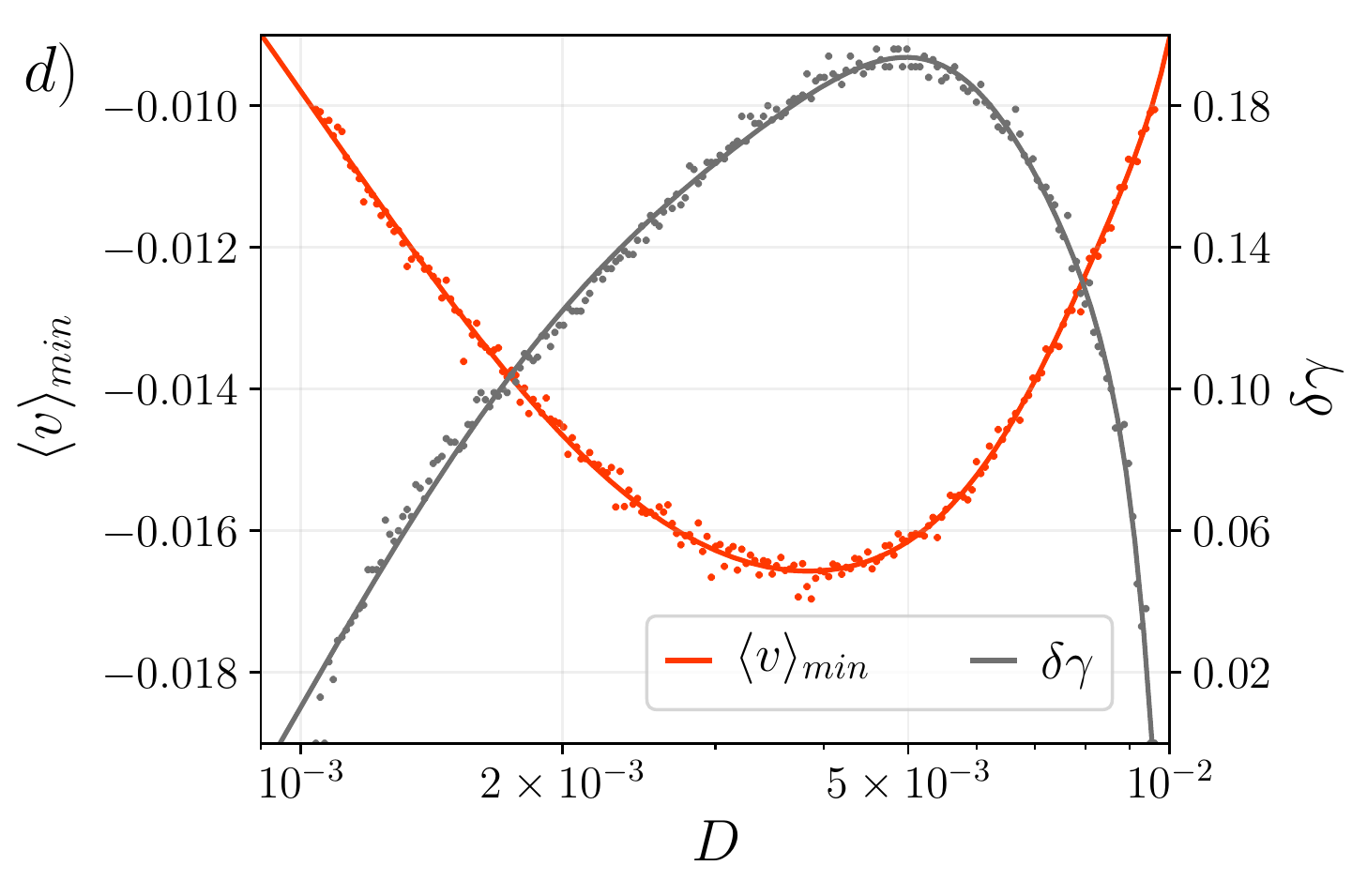} 	  
	\caption{Panel (a)-(c): The friction coefficient (proportional to the particle size) $\gamma^* \propto R$ for which the Brownian particle velocity attains its global minimum $\langle v \rangle_{min} \equiv \langle v \rangle(\gamma^*)$ as a function of temperature $D \propto T$. The grey bars represent the friction coefficient interval $\delta \gamma$ in which the ANM effect takes place (c.f. Fig. \ref{fig3} (b)). The cyan region marks the range of the particle size $\gamma \propto R$ corresponding to the vicinity of minimum, i.e. $[\langle v \rangle_{min} - 0.05\langle v \rangle_{min}, \langle v \rangle_{min}]$. Other parameters read (a): $a = 5.55$, $\omega = 1.5$; (b): $a = 5.75$, $\omega = 1.65$; (c): $a = 5.75$, $\omega = 5.65$. In all panels the bias $f = 0.1$. Plot (d) depicts the minimal velocity $\langle v \rangle_{min}$ (left axis) as well as the resolution capacity $\delta \gamma$ (right axis) versus temperature $D$ for the parameter regime corresponding to the panel (c).}
	\label{fig4}
\end{figure*}

In panel (a) of Fig. \ref{fig1} we demonstrate the ANM phenomenon. For $\gamma = 2.135$ the average velocity $\langle v \rangle$ assumes the same sign as the force $f$, which results in typical transport behaviour, consistent with the direction of the applied bias. For $\gamma = 1.935$, however, the average response of the system is opposite to the acting static bias $f$, i.e., $\langle v \rangle < 0$ for small $f > 0$, indicating the ANM phenomenon. There is no relationship detectable between the set of parameter values and the occurrence of the ANM effect and therefore the values of the friction coefficient chosen here are just exemplary ones. It has been shown that there exist two fundamentally different mechanisms responsible for the emergence of ANM in our setup: (i) it can be generated by the deterministic chaotic dynamics \cite{speer2007} or (ii) be induced by thermal equilibrium fluctuations \cite{machura2007}. 
Very recently, a third mechanism generating the ANM effect has been discovered. Accordingly, ANM phenomenon may emerge as well within deterministic and non-chaotic parameter regimes \cite{slapik2018}.

In panel (b) of the Fig. \ref{fig1} we study temperature dependence of the average velocity $\langle v \rangle$ in the exemplary parameter regime corresponding to the ANM effect. For $\gamma = 1.895$ the ANM phenomenon is observed even in the limit of vanishing thermal noise intensity $D \propto T \to 0$ indicating that in this regime the latter effect is caused purely by the deterministic dynamics of the system. This mechanism is the most populated in the parameter space. As it is illustrated in the panel, in such a case thermal fluctuations typically have destructive impact, i.e., when thermal noise intensity is increasing the ANM disappears. On the other hand, for $\gamma = 1.935$ the ANM is induced by thermal fluctuations. It means that there is a finite window of temperature $D \propto T$ in which the average velocity $\langle v \rangle < 0$ with $f > 0$ and this effect is not observed in the limit of vanishing thermal noise intensity $D \to 0$. 

We now attempt to find a parameter regime for which particles differing by size could be isolated by appropriate dose of thermal fluctuations $D \propto T$. Motivated by a large size range typically encountered in biochemical applications we aim to develop a tunable scheme which allows to control particle size targeted for isolation by changing solely temperature of the system. In doing so we are automatically restricted only to parameter sets corresponding to thermal noise induced ANM as for the deterministic mechanism temperature has destructive influence on the latter effect. These regimes are significantly less populated in the parameter space than the deterministic ones and therefore this task is highly challenging.

As the first step of our analysis we isolated all parameter regimes $\{a, \omega, f, D\}$  for which the ANM effect is induced by thermal fluctuations and observed for only one narrow interval of the friction coefficient $\gamma$. We exemplify this procedure in Fig. \ref{fig3} where we depict the average velocity $\langle v \rangle$ versus the friction $\gamma$ which can be identified with the particle size $R$. Amid many particles of sizes corresponding to the friction in a wide interval $\gamma \in [10^{-1},10]$ only those with coefficient $\gamma^* = 1.59$ will move in the opposite direction $\langle v \rangle < 0$ to the applied bias $f > 0$. Other particles will follow towards the direction of the bias. As a consequence, only those particles with $\gamma^* \approx 1.59$ will be separated from the others. In the panel (b) we magnify the interval in which the ANM phenomenon occurs. It can be viewed as a resolution capacity of this method. In the presented case it reads $\delta \gamma \approx 0.1$. However, as it is illustrated, typically negative velocity $\langle v \rangle$ is noticeably peaked in such a region and therefore the particles of the precisely defined size $\gamma^*$ for which the velocity is minimal $\langle v \rangle_{min} = \langle v \rangle(\gamma^*)$ will be pronouncedly isolated from the others. We stress that the dimensionless friction coefficient $\gamma$ in Eq. (\ref{gamma}) depends not only on the actual friction $\Gamma \propto R$ but also on the parameters of the potential $\Delta U$ and $L$. Therefore experimentalists may exploit these characteristics of the periodic substrate to further adapt the particle size targeted for separation.

Given the target audience of this journal we now provide the exemplary set of the model parameters expressed in real, physical units corresponding to the regime presented in Fig. \ref{fig3}. In doing so we harvest the data for the microfluidic system which in Ref. \cite{luo2016} was successfully exploited to separate colloidal particles and mouse liver mitochondria by utilizing the ANM effect. In particular, for a realistic colloidal particle of the radius \mbox{$R = 2.2 \, \mu \mathrm{m}$}, for which the negative mobility effect is tailored, suspended in aqueous solution with the viscosity $\eta = 8.9 \cdot 10^{-4} \,\mathrm{Pa}\,\mathrm{s}$ at temperature $T = 25\,^{\circ} \mathrm{C}$, the characteristic time scales are $\tau_{\gamma} = 0.4\, \mathrm{s}$ and $\tau_0 = 0.25 \, \mathrm{s}$. These imply that the potential barrier $\Delta U = 43 \, \mathrm{eV}$ and the spatial period of the potential $L = 11 \, \mu \mathrm{m}$. Consequently, the conservative force $-U'(x)$ is of the order $\Delta U/L = 0.63 \, \mathrm{pN}$. The amplitude and the frequency of the external driving reads \mbox{$A = 3.6 \, \textrm{pN}$} and $\Omega = 10 \,\mathrm{Hz}$, respectively. Finally, the constant bias is $F = 0.063 \, \mathrm{pN}$. We note that all characteristic time scales $\tau_{\gamma} = 0.4 \, \mathrm{s}$, $\tau_0 = 0.25 \, \mathrm{s}$ and \mbox{$\mathsf{T} = 2\pi/\Omega = 0.628 \, \mathrm{s}$} are of the same order of magnitude which is typical for the parameter regimes in which the ANM effect is observed. Moreover, the order of magnitude of $\mathrm{pN}$ appearing here is adequate for the biomolecular scale as e.g. the Brownian motion force on an \emph{E.coli} bacterium averaged over 1 second is $0.01 \, \mathrm{pN}$ and the propulsion developed by a molecular motor is $5 \, \mathrm{pN}$ \cite{howard}.

As the second step of the analysis we focused on the parameter regimes $\{a, \omega, f\}$ for which a specific functional dependence between the friction $\gamma^*$ (size of the particle) intended for isolation and temperature $D \propto T$ can be revealed. In Fig. \ref{fig4} (a)-(c) we present three \emph{exemplary} curves $\gamma^*(D)$ for different values of the external harmonic driving amplitude $a$ and the frequency $\omega$. They have been obtained from characteristics $\langle v \rangle(\gamma)$ computed for many different temperatures $D \propto T$, c.f. Fig. \ref{fig3} (a). Each blue dot in the plot represents the friction coefficient $\gamma^*$ for which the Brownian particle velocity attains its global minimum value at fixed thermal fluctuations intensity. The grey bars indicate the friction coefficient interval $\delta \gamma$ where the ANM effect occurs, c.f. Fig. \ref{fig3} (b). The conclusion is that there is no single parameter regime covering a wide range of the particle size $\gamma^*$ targeted for separation by changing solely temperature $D \propto T$. However, the parameters $a$, $\omega$ and $f$ give enough freedom to cover by parts a physically significant interval of moderate to large friction which is characteristic for small particles operating at low Reynolds numbers, e.g. for the panel (a) $\gamma^* \in [0.84,0.96]$; (b) $\gamma^* \in [0.55,0.58]$; (c) $\gamma^* \in [0.58,0.75]$. Using one of these exemplary tailored parameter regimes one is able to tune the ANM to the particle of a given size $\gamma^*$ by changing solely temperature $D \propto T$ of the system. Doing so allows to separate particles with respect to their size in an efficient and tunable way.
\section{Discussion}
At first glance the magnitude of intervals $\delta \gamma$ where the ANM occurs which are represented in panels Fig. \ref{fig4} (a)-(c) by the grey bars may look alarming. However, this fact should be considered as an \emph{intrinsic feature} rather than a bug. First, let us discuss the typical dependence of the resolution capacity $\delta \gamma$ on temperature $D$. This function is depicted in the panel (d) (right axis) for the parameter regime corresponding to the panel (c). The reader may observe there the non-monotonic dependence of $\delta \gamma$ on temperature $D$ which is very characteristic for the negative mobility phenomenon induced by thermal fluctuations. The anomalous transport effect emerges at the minimal temperature $D_{min}$, then $\delta \gamma$ initially grows, passes through its maximum observed for an appropriate dose of thermal noise and then starts to decrease until it reaches the maximal temperature $D_{max}$. In the same panel we depict also the minimal velocity $\langle v \rangle_{min} = \langle v \rangle(\gamma^*)$, corresponding to the particle size $\gamma^*$ for which the negative mobility is tailored, on temperature $D$. Luckily, we observe that initially thermal noise not only increases the interval $\delta \gamma$ where the ANM is detected but at the same time it enhances the absolute value of the global minimum of the Brownian particle velocity $\langle v \rangle_{min}$. Moreover, we can notice that there is temperature for which the resolution capacity $\delta \gamma$ is maximal and the velocity $\langle v \rangle_{min}$ is minimal. It means that even though as thermal noise increases the ANM peaks in characteristics $\langle v \rangle(\gamma)$ become wider they also get more pronounced, c.f. Fig. \ref{fig3} (b). This fact guarantees that the particle size $\gamma^*$ for which the ANM is tailored will be well distinguished from the others. For this reason in the panel (a)-(c) of Fig. \ref{fig4} we additionally marked with the cyan color the region of the particle size $\gamma$ corresponding to the vicinity of minimum, i.e. $[\langle v \rangle_{min} - 0.05\langle v \rangle_{min}, \langle v \rangle_{min}]$. In this range $\delta \gamma$ typically equals several percent of the value $\gamma^*$ which is reasonable for the separation purposes. Note that since this interval is a projection of the window $[\langle v \rangle_{min} - 0.05\langle v \rangle_{min}, \langle v \rangle_{min}]$ onto $\gamma$-axis, the so obtained bar $\delta \gamma$ is not necessarily symmetric around $\gamma^*$.

Second, taking into account the above discussion, let us point out that even though there will be two particle species differing by size $\gamma_1 \propto R_1$ and $\gamma_2 \propto R_2$ both in the range where ANM occurs, c.f. grey area in Fig. \ref{fig3} (b), \emph{typically} they will be still separated from each other since usually $\langle v \rangle(\gamma_1) \neq \langle v \rangle(\gamma_2)$. It means that, unfortunately, they both will travel into the direction opposite to the applied constant force $f$ but there will be a gap between them because of the difference in their velocities. This opens up an intriguing possibility for simultaneous separation of several particle species under the identical experimental conditions.

Let us now comment on the salient difference between this work and our previous paper \cite{slapik2019}. The obvious one is that here we discuss the separation scheme with respect to the particle size $\gamma \propto R$ rather than its mass $m$. Moreover, in our previous paper we harvested the negative mobility effect whose roots lied solely in the deterministic dynamics of the system, c.f. Fig. \ref{fig1} (b). As the latter mechanism is significantly more populated in the parameter space of the model than the thermal noise induced one it allowed us to find an unique parameter set for which the particle mass intended for separation is effectively controlled over a regime of nearly two orders of mass magnitude upon changing solely the frequency $\omega$ of the external harmonic driving. Another difference is that typically the deterministic ANM is quickly destroyed by temperature growth meaning that the mentioned parameter regime works effectively only for low thermal noise intensities. As a consequence the ANM peaks in the characteristics $\langle v \rangle(m)$ are much steeper, resolution capacity $\delta m$ is significantly better and the overall selectivity of the method is superior. On the other hand, here the negative mobility effect is induced by thermal fluctuations
meaning that one is able to find parameter set allowing to observe the ANM in the temperature regime as high as $D = 10^{-2}$, c.f. Fig. \ref{fig4} (c), which is not the case for the deterministic mechanism. Moreover, only due to this different origin the negative mobility effect can be controlled by thermal fluctuations. This fact combined with an appropriate experimental implementation opens an opportunity to separate particles that carry no charge or dipole or they can hardly be manipulated by means of an external field or force. The prize that needs to be paid for this possibility is a bit loss of accuracy in the separation process.

Last but not least, let us make a short comment on the hydrodynamic corrections that may play a key role in experimental reality. We have taken into account only the simplest hydrodynamic effect expressed by the Stokes term in the model equation (\ref{model}) and neglected a number of additional phenomena which may prove experimentally appreciable. In particular, when the particle travels in a system with geometrical constraints, which is typically the case e.g. in the microfluidic device, its boundaries significantly modify the particle dynamics. The geometry usually increases the hydrodynamic drag. This effect is notoriously difficult to treat analytically and numerically \cite{happel1965} and certainly lies beyond the scope of this paper. Nevertheless for some systems it appears to be extremely important and may cause significant underestimation of the results \cite{yang2017}. It can be accurately incorporated by a phenomenological modification based on the experimentally measured quantities and appropriate rescaling of the model parameters \cite{yang2017}. Since the hydrodynamic drag is increased for such systems we expect that this effect is likely to hamper the direct observation of the ANM phenomenon. Our preliminary results tell that the latter anomalous transport effect notably decreases with the increase of friction (drag) (not depicted). Moreover, simultaneously the ANM becomes less populated in the parameter space what makes it even harder to detect (not shown). Therefore we are aware of the fact that our theoretical predictions following from Eq. (\ref{dimlessmodel}) should be used as a guide towards "physical reality" pointing the direction for future experimental and theoretical research rather than taken as granted without approximations.

\section{Conclusion}
In conclusion, this work provides an effective method for the tunable size-based particle separation. In this scheme particle size  intended for isolation can be controlled by changing solely temperature of the system without modifying
the setup. It requires a symmetric spatially periodic nonlinear structure,  an external time periodic driving and a constant bias. 
Our approach can be readily realized using a lab-on-a-chip device \cite{luo2016} by harvesting current lithographic techniques to develop robust separation applications. These may be further enhanced by advances in 3D printing technologies which recently have been exploited down to the nanometer range \cite{beauchamp2017}. For this reason we envision that the investigated method can be adapted to a wide range of separation problems in which size selectivity is required hopefully leading to e.g. new diagnostic applications with a commercial potential \cite{chiu2017}.
\section*{Acknowledgement}
This work has been supported by the Grant NCN No. 2017/26/D/ST2/00543 (J. S.)


\begin{thebibliography}{99}
	\bibitem{yager2006} P. Yager, T. Edwards, E. Fu, K. Helton, K. Nelson, Microfluidic diagnostic technologies for global public health. \textit{Nature} \textbf{442}, 412 (2004)
	\bibitem{cheng2007} X. Cheng, D. Irimia, M. Dixon, K. Sekine, U. Demirci, L. Zamir, R. G. Tompkins, W. Rodriguez, M. Toner, A microfluidic device for practical label-free CD4(+) T cell counting of HIV-infected subjects. \textit{Lab Chip} \textbf{7}, 170 (2007)
	\bibitem{korecka2007} J. A. Korecka, J. Verhaagen, E. M. Hol, Cell-replacement and gene-therapy strategies for Parkinson's and Alzheimer's disease. \textit{Regen. Med.} \textbf{2}, 425 (2007)
	\bibitem{heffner1978} R. R. Heffner, S. A. Barron, The early effects of ischemia upon skeletal muscle mitochondria. \textit{J. Neurol. Sci.} \textbf{38}, 295 (1978)
	\bibitem{eguchi1987} M. Eguchi, Y. Iwama, F. Ochiai, K. Ishikawa, H. Sakakibara, H. Sakamaki, T. Furukawa, Giant mitochondria in acute lymphocytic leukemia. \textit{Exp. Mol. Pathol.} \textbf{47}, 69 (1987)
	\bibitem{suresh2007} S, Suresh. Biomechanics and biophysics of cancer cells. \textit{Acta Mater.} \textbf{55}, 3989 (2007)
	\bibitem{bhagat2010} A. A. Bhagat, H. Bow, S. W. Hou, S. J. Tan, J. Han, C. T. Lim, Microfluidics for cell separation. \textit{Med. Biol. Eng. Comput.} \textbf{48}, 999 (2010)
	\bibitem{xuan2014} J. Xuan, M. L. Lee, Size separation of biomolecules and bioparticles using micro/nanofabricated structures. \textit{Anal. Methods} \textbf{6}, 27 (2014)
	\bibitem{sajeesh2014} P. Sajeesh, A. K. Sen, Particle separation and sorting in microfluidic devices: a review. \textit{Microfluid. Nanofluid.} \textbf{17}, 1 (2014)
	\bibitem{sonker2019} M. Sonker, D. Kim, A. Egatz-Gomez, A. Ros, Separation Phenomena in Tailored Micro- and Nanofluidic Environments. \textit{Annu. Rev. Anal. Chem.} \textbf{12}, 475 (2019)
	\bibitem{eichhorn2002} R. Eichhorn, P. Reimann, P. H\"anggi, Brownian motion exhibiting absolute negative mobility. \textit{Phys. Rev. Lett.} \textbf{88}, 190601 (2002)
	\bibitem{machura2007} {\L}. Machura, M. Kostur, P. Talkner, J. {\L}uczka, P. H\"anggi, Absolute negative mobility induced by thermal equilibrium fluctuations. \textit{Phys. Rev. Lett.} \textbf{98}, 040601 (2007)
	\bibitem{spiechowicz2014pre} J. Spiechowicz, P. H\"anggi, J.{\L}uczka, Brownian motors in the microscale domain: Enhancement of efficiency by noise. \textit{Phys. Rev. E} \textbf{90}, 032104 (2014)
	\bibitem{slapik2019} A. Slapik, J. {\L}uczka, P. H\"anggi, J. Spiechowicz, Tunable Mass Separation via Negative Mobility. \textit{Phys. Rev. Lett.} \textbf{122}, 070602 (2019)	
	\bibitem{ros2005} A. Ros, R. Eichhorn, J. Regtmeier, T. T. Duong, P. Reimann, D.Anselmetti, Absolute negative mobility. \textit{Nature} \textbf{436}, 928 (2005)
	\bibitem{eichhorn2010} R. Eichhorn, J. Regtmeier, D. Anselmetti, P. Reimann, Negative mobility and sorting of colloidal particles. \textit{Soft Matter} \textbf{6}, 1858 (2010)
	\bibitem{luo2016} J. Luo, K. Muratore, E. Arriaga, A. Ros, Deterministic Absolute Negative Mobility for Micro- and Submicrometer Particles Induced in a Microfluidic Device. \textit{Anal. Chem.} \textbf{88}, 5920 (2016)
	\bibitem{hanggi2009} P. H\"anggi, F. Marchesoni, Artificial Brownian motors: Controlling transport on the nanoscale. \textit{Rev. Mod. Phys.} \textbf{81}, 387 (2009)
	\bibitem{spiechowicz2016jstatmech} J. Spiechowicz, J. {\L}uczka, {\L}. Machura, Efficiency of transport in periodic potentials: dichotomous noise contra deterministic force. \textit{J. Stat. Mech}, 054038 (2016)
	\bibitem{spiechowicz2017scirep} J. Spiechowicz, J. {\L}uczka, Subdiffusion via dynamical localization induced by thermal equilibrium fluctuations. \textit{Sci. Rep.} \textbf{7}, 16451 (2017)
	\bibitem{reimann2001} P. Reimann, C. Van den Broeck, H. Linke, P. H\"anggi, J. M. Rubi, A. Perez-Madrid, Giant acceleration of free diffusion by use of tilted periodic potentials. \textit{Phys. Rev. Lett.} \textbf{87}, 010602 (2001)
	\bibitem{spiechowicz2015chaos} J. Spiechowicz, J. {\L}uczka, Josephson phase diffusion in the superconducting quantum interference device ratchet. \textit{Chaos} \textbf{25}, 053110 (2015)
	\bibitem{lindner2016} B. Lindner, I. M. Sokolov, Giant diffusion of underdamped particles in a biased periodic potential. \textit{Phys. Rev. E} \textbf{93}, 042106 (2016)
	\bibitem{spiechowicz2016njp} J. Spiechowicz, P. Talkner, P. H\"anggi, J. {\L}uczka, Non-monotonic temperature dependence of chaos-assisted diffusion in driven periodic systems. \textit{New J. Phys.} \textbf{18}, 123029 (2016)
	\bibitem{spiechowicz2017chaos} J. Spiechowicz, M. Kostur, J. {\L}uczka, Brownian ratchets: How stronger thermal noise can reduce diffusion. \textit{Chaos} \textbf{27}, 023111 (2017)
	\bibitem{spiechowicz2016scirep} J. Spiechowicz, J. {\L}uczka, P. H\"anggi, Transient anomalous diffusion in periodic systems: ergodicity, symmetry breaking and velocity relaxation. \textit{Sci. Rep.} \textbf{6}, 30948 (2016)
	\bibitem{spiechowicz2015cpc} J. Spiechowicz, M. Kostur, {\L}. Machura, GPU accelerated Monte Carlo simulation of Brownian motors dynamics with CUDA. \textit{Comp. Phys. Commun.} \textbf{191}, 140 (2015)
	\bibitem{platen} E. Platen and N. Bruti-Liberati, \textit{Numerical Solution of Stochastic Differential equations with Jumps in Finance (Stochastic Modelling and Applied Probability)}, (Springer, Berlin 2010)
	\bibitem{speer2007} D. Speer, R. Eichhorn, P. Reimann, Transient chaos induces anomalous transport properties of an underdamped Brownian particle. \textit{Phys. Rev. E} \textbf{76}, 051110 (2007)
	\bibitem{nagel2008} J. Nagel, D. Speer, T. Gaber, A. Sterck, R. Eichhorn, P. Reimann, K. Ilin, M. Siegel, D. Koelle, R. Kleiner, Observation of Negative Absolute Resistance in a Josephson Junction. \textit{Phys. Rev. Lett.} \textbf{100}, 217001 (2008)
	\bibitem{slapik2018} A. Slapik, J. {\L}uczka, J. Spiechowicz, Negative mobility of a Brownian particle: Strong damping regime. \textit{Commun. Nonlinear Sci. Numer. Simul.} \textbf{55}, 316 (2018)
	\bibitem{howard} J. Howard, Mechanics of Motor Proteins and the Cytoskeleton (Sinauer Associates, Massachusetts, USA, 2001)
	\bibitem{happel1965} J. Happel and H. Brenner, Low Reynolds Number Hydrodynamics (Prentice Hall, Englewood Cliffs, NJ, 1965)
	\bibitem{yang2017} X. Yang, C. Liu, Y. Li, F. Marchesoni, P. H\"anggi and H. P. Zhang, Hydrodynamic and entropic effects on colloidal diffusion in corrugated channels, Proc. Natl. Acad. Sci. 114, 9564–9569 (2017)
	\bibitem{beauchamp2017} M. J. Beauchamp, G. P. Nordin, A. T. Woolley, Moving from millifluidic to truly microfluidic sub-100-$\mu$m cross-section 3D printed devices. \textit{Anal. Bioanal. Chem.} \textbf{409}, 11 (2017)
	\bibitem{chiu2017} D. T. Chiu, D. Di Carlo, P. S. Doyle, C. Hansen, R. M. Maceiczyk, R. C. Wootton, Small but perfectly formed? Successes, challenges and opportunities for microfluidics in the chemical and biological sciences. \textit{Chemistry} \textbf{2}, 201 (2017)
\end{thebibliography}
\end{document}